\documentclass[english,showpacs,floatfix,preprint,10pt]{revtex4}
\usepackage[dvips]{graphicx}
\usepackage{longtable}
\usepackage{amsmath,amssymb}
\usepackage{dcolumn}
\usepackage{latexsym}
\usepackage{babel}
\usepackage[latin1]{inputenc}
\usepackage{color}
\usepackage[colorlinks]{hyperref}

\begin{document}
\baselineskip=0.8 cm
\title{{\bf Effects of dark sectors' mutual interaction on the growth of structures}}

\author{Jian-Hua He$^{1}$, Bin Wang$^{1}%
$\footnote{wangb@fudan.edu.cn }, Y. P.
Jing$^{2}$\footnote{ypjing@shao.ac.cn}} \affiliation{$^{1}$
Department of Physics, Fudan University, 200433 Shanghai, China}
\affiliation{$^{2}$ Shanghai Astronomical Observatory, Nandan Road
80, Shanghai, 200030, China}

\vspace*{0.2cm}
\begin{abstract}
\baselineskip=0.6 cm

We present a general formalism to study the growth of dark matter
perturbations when dark energy perturbations and interactions
between dark sectors are present. We show that the dynamical
stability on the growth of structure depends on the form of coupling
between dark sectors. By taking the appropriate coupling which
enables the stable growth of structure, we find that the effect of
the interaction between dark sectors overwhelms that of dark energy
perturbation on the growth of dark matter perturbation. Due to the
influence of the interaction, the growth index can differ from the
value without interaction by an amount up to the observational
sensibility, which provides an opportunity to probe the interaction
between dark sectors through future observations on the growth of
structure.

\end{abstract}

 \maketitle
\newpage
\section{Introduction}

There has been growing observational evidence indicating that our
universe has entered the epoch of accelerated expansion \cite{1,2}.
Within the framework of Einstein gravity, this acceleration can be
attributed to the so called dark energy (DE) with negative pressure,
which dominates the content of universe at present. The leading
interpretation of such DE is a cosmological constant with equation
of state (EoS) $w =-1$ \cite{3}. Though the cosmological constant is
consistent with the observation data, at the fundamental level it
fails to be convinced: the vacuum energy density falls far below the
value predicted by any sensible quantum field theory by many orders
of magnitude, and it unavoidably leads to the coincidence problem,
i.e., ``why are the vacuum and matter energy densities of precisely
the same order today?". More sophisticated models have been proposed
to replace the cosmological constant by a dynamical dark energy in
the conjectures relating the DE either to a scalar field called
quintessence with $w>-1$, or to an exotic field called phantom with
$w <-1$. It would be fair to say that there is no clear winner in
sight to explain the nature of DE at the moment. On the other hand,
the remarkable discovery of the cosmic acceleration might also be
doubted as the result of a modification of standard Einstein gravity
at large distances. This happens in $f(R)$ theories \cite{4} and in
braneworld models\cite{5}. How to distinguish such modified gravity
theory from that of DE is an imperative task which will bring about
not only a breakthrough in cosmology, but also in the field of high
energy physics.

The DE is usually not supposed to clump on the scales of the largest
cosmic structures, and the most powerful way where its nature can be
unveiled is to investigate the expansion history of
universe\cite{6}. However, the current observations of cosmic
expansion history cannot break the degeneracies among different
approaches trying to explain the cosmic acceleration. Considering
that the rate of expansion as a function of cosmic time can in turn
affect the process of structure formation, an alternative channel to
study the clustering properties of cosmic structure has been used to
detect the possible effects of DE. It is expected that the growth
history of dark matter (DM) perturbations can provide a
complementary probe to distinguish the modified gravity from
DE\cite{7}. However, most of these studies have neglected the effect
of DE perturbations. The DE perturbations do not affect the
background evolution, but they are crucial in determining the DE
clustering properties\cite{zhang}, which consequently have effects
on the evolution of DM perturbations. Recently it has been found
that for small sound speed and DE EoS prominently away from $-1$,
the presence of DE perturbations may leave a significant imprint on
the growth function of DM perturbations\cite{8}.

In the minimal picture, DM does not feel any significant
interactions from DE. Although this picture is consistent with
current observations, considering that DE accounts for a large
fraction of the universe, it is natural, in the framework of field
theory, to consider its interaction with the remaining fields in the
universe. The possibility that DE and DM can interact has been
widely discussed in \cite{9}-\cite{31}. It has been shown that the
coupling between DE and DM can provide a mechanism to alleviate the
coincidence problem \cite{9,91,93,99}. In addition, it has been
argued that an appropriate interaction between DE and DM can
influence the perturbation dynamics and affect the lowest multipoles
of the CMB angular power spectrum \cite{10,11}. Furthermore it was
suggested that the dynamical equilibrium of collapsed structures
such as clusters would be modified due to the coupling between DE
and DM \cite{13,22}. In the presence of such coupling, there has
been some concerns about the stability of the perturbations
\cite{maartens}. However, it was proved in \cite{he} that the
stability of the curvature perturbation depends on the forms of
coupling between dark sectors. Since observational signatures on the
dark sectors' mutual interaction have been found in the probes of
the cosmic expansion history \cite{16,23,24}, it is interesting to
ask whether the interaction can have an effect on the growth of
structure, whether it can provide a more consistent check for the
coupling between dark sectors.

In this paper we are going to study the effect of the interaction
between DE and DM on the growth function of DM perturbations.
Recently there is an attempt on this study\cite{new}. Here we will
incorporate the DE perturbations in our study. We will restrict our
investigation to constant sound speed as well as constant EoS of DE.
From our formalism we will see that the growth of DM perturbations
gets more modification due to the dark sectors' interaction than
that of the DE perturbations. This provides an opportunity to probe
the interaction between DE and DM through the growth history of DM.

The organization of the paper is as follows. In the following
section we will provide our analytical framework for the
perturbation equations at the linear level. In Sec.III, we will
present our numerical results and discuss the growth function of
DM. We will give our conclusions and discussions in the last
section.

\section{Analytical formalism}
In this section we will derive the second order differential
equations for the perturbations of DM and DE with couplings between
them in  a spatially flat Friedmann-Robertson-Walker (FRW)
background. The perturbed space-time at first order reads,
\begin{equation}
ds^2 = a^2[-(1+2\psi)d\tau^2+2\partial_iBd\tau
dx^i+(1+2\phi)\delta_{ij}dx^idx^j+D_{ij}Edx^idx^j],\label{perturbedspacetime}
\end{equation}
where $\psi, B,\phi, E$ is scalar metric perturbations, $a$ is the
cosmic scale factor and $
D_{ij}=(\partial_i\partial_j-\frac{1}{3}\delta_{ij}\nabla^2)$.

We work with general stress-energy tensor
\begin{equation}
T^{\mu\nu}=\rho U^{\mu}U^{\nu}+p(g^{\mu\nu}+U^{\mu}U^{\nu}),
\end{equation}
for a two-component system consisting of DE and DM. Each
energy-mementum tensor satisfies the conservation law
\begin{eqnarray}
\nabla_{\mu}T^{\mu\nu}_{(\lambda)}=Q^{\nu}_{(\lambda)}
\end{eqnarray}
where $Q^{\nu}_{(\lambda)}$ denotes the interaction between
different components and $\lambda$ denotes either the DM or the DE
sector. The perturbed energy-momentum tensor reads,
\begin{eqnarray}
\delta \nabla_{\mu}T^{\mu0}_{(\lambda)} &=&
\frac{1}{a^2}\{-2[\rho_{\lambda}'+3\mathcal
{H}(p_{\lambda}+\rho_{\lambda})]\psi+\delta
\rho_{\lambda}'+(p_{\lambda}+\rho_{\lambda})\theta_{\lambda}+3\mathcal
{H}(\delta p_{\lambda}+\delta
\rho_{\lambda})+3(p_{\lambda}+\rho_{\lambda})\phi'\}\nonumber\\
&=&\delta Q^0_{\lambda}\nonumber \label{perturbation}
\\
\partial_i \delta \nabla_{\mu}T^{\mu i}_{(\lambda)} &=&
\frac{1}{a^2}\{[p'_{\lambda}+\mathcal
{H}(p_{\lambda}+\rho_{\lambda})]\nabla^2B+[(p'_{\lambda}+\rho'_{\lambda})+4\mathcal{H}(p_{\lambda}+\rho_{\lambda})]\theta_{\lambda}\\\nonumber
&&+(p_{\lambda}+\rho_{\lambda})\nabla^2B'+\nabla^2\delta
p_{\lambda}+(p_{\lambda}+\rho_{\lambda})\theta_{\lambda}'+(p_{\lambda}+\rho_{\lambda})\nabla^2\psi\}
= \partial_i \delta Q^i_{(\lambda)}
\end{eqnarray}
where $\theta = \nabla^2 v$, $v$ is the potential of three velocity
and the prime denotes the derivative with respect to the conformal
time $\tau$. The perturbed Einstein equations yield, in the linear
order
\begin{eqnarray}
\nabla^2\phi+3\mathcal{H}\left(\mathcal{H}\psi-\phi'\right)+\mathcal{H}\nabla^2B-\frac{1}{6}[\nabla^2]^2E&=&-4\pi
Ga^2\delta \rho\nonumber \\
\mathcal{H}\nabla^2\psi-\nabla^2\phi'+2\mathcal{H}^2\nabla^2B-\frac{a^{''}}{a}\nabla^2B+\frac{1}{6}[\nabla^2]^2E'&=&-4\pi
G a^2(\rho+p)\theta\nonumber \label{PerEinstein}\\
-\partial^{i}\partial_{j}\psi-\partial^{i}\partial_{j}\phi+\frac{1}{2}\partial^{i}\partial_{j}E^{''}+\mathcal{H}\partial^{i}\partial_{j}E'+\frac{1}{6}\partial^{i}\partial_{j}\nabla^2E-2\mathcal{H}\partial^{i}\partial_{j}B-\partial^{i}\partial_{j}B'&=&8\pi
Ga^2\Pi^{i}_{j}
\end{eqnarray}
where $\delta \rho$ is the total energy perturbation, $\delta \rho =
\sum_{\lambda} \delta \rho_{\lambda}$ and $(p+\rho)\theta=
\sum_{\lambda}(p_{\lambda}+\rho_{\lambda})\theta_{\lambda}$.

Considering an infinitesimal transformation on the
coordinates\cite{Sa},
\begin{eqnarray}
\tilde{x}^{\mu}&=& x^{\mu}+\delta x^{\mu}\nonumber \\
\delta x^0 &=& \xi^0(x^{\mu})\nonumber\\
\delta x^{i} &=& \partial^{i}\beta
(x^{\mu})+v_*^{i}(x^{\mu}) (\partial_iv_*^{i}=0)\nonumber\\
\end{eqnarray}
the perturbed quantities behave as,
\begin{eqnarray}
\tilde{\psi} &=& \psi -\xi^{0'}-\frac{a'}{a}\xi^0\nonumber\\
\tilde{B} &=& B +\xi^{0}-\beta'\nonumber\\
\tilde{\phi} &=& \phi
-\frac{1}{3}\nabla^2\beta-\frac{a'}{a}\xi^0\nonumber\\
\tilde{E}&=&E-2\beta\nonumber\\
\tilde{v}&=&v+\beta'\nonumber \\\label{quantities}
\tilde{\theta}&=&\theta+\nabla^2\beta'.
\end{eqnarray}
Inserting eq(\ref{quantities})into the medium  part of
eq(\ref{perturbation}), we obtain the behavior of $\delta Q^{\mu}$
on the right hand side of eq(\ref{perturbation})
\begin{eqnarray}
\tilde{\delta Q^0}&=&\delta Q^0-Q^{0'}\xi^0+Q^0\xi^{0'}\nonumber\\
\tilde{\delta Q_p}&=&\delta Q_p+Q^0 \beta',\label{Qtransform}
\end{eqnarray}
where
\begin{equation}
\delta Q^i= \partial^i\delta Q_p+ \delta Q_*^i(\partial_i\delta
Q_*^i=0).\nonumber
\end{equation}
and $\delta Q_p$ is the potential of three vector $\delta Q^i$. This
is consistent with the results got from Lie derivatives,
\begin{eqnarray}
\mathcal{L}_{\delta x} Q^{\nu} &=& \delta
x^{\sigma}Q^{\nu}_{,\sigma}-Q^{\sigma}\delta
x^{\nu}_{,\sigma}\label{Lie}\nonumber\\
\delta \tilde{Q}^{\nu}&=&\delta Q^{\nu}-\mathcal{L}_{\delta x}
Q^{\nu},
\end{eqnarray}
which shows that $Q^{\nu}$ is covariant.

We expand the metric perturbations in Fourier space by using
scalar harmonics\cite{Sa},
\begin{eqnarray}
\tilde{\psi}Y^{(s)} &=& (\psi -\xi^{0'}-\frac{a'}{a}\xi^0)Y^{(s)}\nonumber\\
\tilde{B}Y_i^{(s)} &=& (B -k\xi^{0}-\beta')Y_i^{(s)}\nonumber\\
\tilde{\phi}Y^{(s)} &=& (\phi
-\frac{1}{3}k\beta-\frac{a'}{a}\xi^0)Y^{(s)}\nonumber\\
\tilde{E}Y_{ij}^{(s)}&=&(E+2k\beta)Y_{ij}^{(s)}\nonumber\label{gaugeInfourier}\\
\tilde{\theta}Y^{(s)}&=&(\theta+k\beta')Y^{(s)}
\end{eqnarray}
and the perturbed conservation equations eq(~\ref{perturbation})
read
\begin{eqnarray}
\delta_{\lambda}'+3\mathcal{H}(\frac{\delta p_{\lambda}}{\delta
\rho_{\lambda}}-w_{\lambda})\delta_{\lambda}=
-(1+w_{\lambda})kv_{\lambda}
-3(1+w_{\lambda})\phi'+(2\psi-\delta_{\lambda})\frac{a^2Q^0_{\lambda}}{\rho_{\lambda}}+\frac{a^2
\delta Q^0_{\lambda}}{\rho_{\lambda}}\nonumber\\\label{perturbed}
(v_{\lambda}+B)'+\mathcal{H}(1-3w_{\lambda})(v_{\lambda}+B) =
\frac{k}{1+w_{\lambda}}\frac{\delta p_{\lambda}}{\delta
\rho_{\lambda}}\delta_{\lambda}
-\frac{w_{\lambda}'}{1+w_{\lambda}}(v_{\lambda}+B)+k\psi
-\frac{a^2Q^0_{\lambda}}{\rho_{\lambda}}v_{\lambda}-\frac{w_{\lambda}a^2Q^0_{\lambda}}{(1+w_{\lambda})\rho_{\lambda}}B+\frac{a^2\delta
Q_{p\lambda}}{(1+w_{\lambda})\rho_{\lambda}}.
\end{eqnarray}
The interaction $Q_{(\lambda)}^{\nu}$ can be decomposed into two
parts,
\begin{eqnarray}
Q_{(\lambda)}^{\nu}=Q_{\lambda}U^{\nu}_{(total)}+F_{\lambda}^{\nu}
\end{eqnarray}
where $U_{(total)}^{\mu}$ is the total four-velocity as defined
in\cite{Sa}\cite{maartens},
$F_{\lambda}^{\nu}=h^{\nu}_{\mu}Q_{(\lambda)}^{\mu}$ is a spacial
vector and vanishes in back ground $F_{\lambda}^{\nu}=0$\cite{he},
$h^{\mu \nu}=g^{\mu \nu}+U^{\mu}_{(total)}U^{\nu}_{(total)}$ is the
projection operator. The perturbed quantities read
\begin{eqnarray}
 \delta Q_{\lambda}^0&=&\delta
Q_{\lambda}U^0_{(total)}+Q_{\lambda}\delta U^0_{(total)}\nonumber\\
\delta
Q_{p\lambda}&=&(Q_{\lambda}v_{(total)}+f_{\lambda})U^0_{(total)},
\end{eqnarray}
and Eq(~\ref{perturbed}) can go back to Eq(20) and Eq(21) in
\cite{maartens}, if we neglect the anisotropic stress
$\pi_{\lambda}$.

Constructing the gauge invariant quantities\cite{Sa},
\begin{eqnarray}
\Psi&=& \psi -
\frac{1}{k}\mathcal{H}(B+\frac{E'}{2k})-\frac{1}{k}(B'+\frac{E^{''}}{2k})\nonumber\\
\Phi&=&\phi+\frac{1}{6}E-\frac{1}{k}\mathcal{H}(B+\frac{E'}{2k})\nonumber\\
\delta\rho^I_{\lambda}& = &\delta \rho_{\lambda}-\rho'_{\lambda}\frac{v_{\lambda}+B}{k}\label{gaugeinvariant}\\
\delta p^I_{\lambda}&=& \delta p_{\lambda}-p'_{\lambda}\frac{v_{\lambda}+B}{k}\nonumber\\
\Delta_{\lambda}&=&\delta_{\lambda}-\frac{\rho'_{\lambda}}{\rho_{\lambda}}\frac{v_{\lambda}+B}{k}\nonumber\\
V_{\lambda} &=& v_{\lambda} -\frac{E'}{2k}\nonumber \\
\delta Q^{0I}_{\lambda}&=&\delta
Q^0_{\lambda}-\frac{Q^{0'}_{\lambda}}{\mathcal{H}}(\phi+\frac{E}{6})+Q^{0}_{\lambda}\left[\frac{1}{\mathcal{H}}(\phi+\frac{E}{6})\right]'\nonumber\\
\delta Q_{p\lambda}^{I}&=&\delta Q_{p\lambda }-Q^0_{\lambda
}\frac{E'}{2k}.\nonumber
\end{eqnarray}
the perturbed Einstein equations eq(~\ref{PerEinstein}) become
\begin{eqnarray}
&&\Phi=4\pi G\frac{a^2}{k^2}\sum_{\lambda}\left(\Delta_{\lambda}+\frac{a^2Q_{\lambda}^0}{\rho_{\lambda}}\frac{V_{\lambda}}{k}\right)\rho_{\lambda}\nonumber\\
&&k\left(\mathcal{H}\Psi-\Phi'\right)=4\pi G
a^2\sum_{\lambda}\left(\rho_{\lambda}+p_{\lambda}\right)V_{\lambda}\nonumber \label{Einstein}\\
&&\Psi=-\Phi,
\end{eqnarray}
where we have assumed that the pressure perturbation of DE is
isotropic $\Pi^{i}_j=0$.

Using the gauge invariant quantities eq(~\ref{gaugeinvariant}),  we
can obtain the linear perturbation equations for DM from
eq(\ref{perturbed}),
\begin{eqnarray}
\Delta_m'&+&\left[\frac{\rho_m'}{\rho_m}\frac{V_m}{k}\right]'=-kV_m-3\Phi'+2\Psi\frac{a^2Q_m^0}{\rho_m}-\Delta_m\frac{a^2Q_m^0}{\rho_m}-\frac{\rho_m'}{\rho_m}\frac{V_m}{k}\frac{a^2Q_m^0}{\rho_m}+\frac{a^2\delta
Q^{0I}_m}{\rho_m}\nonumber\\
&&+\frac{a^2Q_m^{0'}}{\rho_m\mathcal{H}}\Phi-\frac{a^2Q_m^0}{\rho_m}\left[\frac{\Phi}{\mathcal{H}}\right]'\label{DM1}\\
V_m'&=&-\mathcal{H}V_m+k\Psi-\frac{a^2Q_m^0}{\rho_m}V_m+\frac{a^2\delta
Q_{pm}^{I}}{\rho_m}\label{DM2}.
\end{eqnarray}
Considering the pressure perturbation of DE \cite{maartens,he}
\begin{equation}
\frac{\delta p_d}{\rho_d} =
C_e^2\delta_d-(C_e^2-C_a^2)\frac{\rho_d'}{\rho_d}\frac{v_d+B}{k},
\end{equation}
where $C_e^2=\frac{\delta p_d}{\delta \rho_d}\mid _{\rm rf}$ is the
sound speed in  DE rest frame , $C_a^2=\frac{p'_d}{\rho'_d}$ is the
adiabatic sound speed , and further noting that $\delta
\tilde{p_d}=\delta p_d -p_d'\xi^0$, we have the gauge invariant form
of DE perturbation equations
\begin{eqnarray}
&&\Delta_d'+\left[\frac{\rho_d'}{\rho_d}\frac{V_d}{k}\right]'+3\mathcal{H}C_e^2(\Delta_d+\frac{\rho_d'}{\rho_d}\frac{V_d}{k})-3\mathcal{H}(C_e^2-C_a^2)\frac{\rho_d'}{\rho_d}\frac{V_d}{k}-3w
\mathcal{H}(\Delta_d+\frac{\rho_d}{\rho_d}\frac{V_d}{k})= \nonumber \\
&&-k(1+w)V_d-3(1+w)\Phi'+2\Psi\frac{a^2Q_d^0}{\rho_d}-(\Delta_d+\frac{\rho_d'}{\rho_d}\frac{V_d}{k})\frac{a^2Q_d^0}{\rho_d}+\frac{a^2\delta
Q^{0I}_d}{\rho_d}+\frac{a^2Q_d^{0'}}{\rho_d\mathcal{H}}\Phi-\frac{a^2Q_d^0}{\rho_d}\left[\frac{\Phi}{\mathcal{H}}\right]'\\
&&V_d'+\mathcal{H}(1-3w)V_d=\frac{k}{1+w}\left[C_e^2(\Delta_d+\frac{\rho_d'}{\rho_d}\frac{V_d}{k})-(C_e^2-C_a^2)\frac{\rho_d'}{\rho_d}\frac{V_d}{k}\right]-\frac{w'}{1+w}V_d+k\Psi-\frac{a^2Q_d^0}{\rho_d}V_d+\frac{a^2\delta
Q_{pd}^{I}}{(1+w)\rho_d}.
\end{eqnarray}

Inserting eq.(\ref{DM2}) into eq.(\ref{DM1})to eliminate $V_m$, in
the subhorizon approximation $k>>aH$, we obtain the second order
equation for the DM perturbation
\begin{equation}
\Delta_m^{''}=-(\mathcal{H}+\frac{2a^2Q_m^0}{\rho_m})\Delta_m'+(-\Delta_m\frac{a^2Q_m^0}{\rho_m}+\frac{a^2\delta
Q^{0I}_m}{\rho_m})(\mathcal{H}+\frac{a^2Q_m^0}{\rho_m})-\Delta_m(\frac{a^2Q_m^0}{\rho_m})'+(\frac{a^2\delta
Q^{0I}_m}{\rho_m})'-\frac{a^2k\delta Q_{pm}^{I}}{\rho_m}-k^2\Psi.
\end{equation}
Similarly for the DE perturbation we have
\begin{eqnarray}
&&\Delta_d^{''}=-3\mathcal{H}'C_e^2\Delta_d-(\frac{a^2Q_d^0}{\rho_d})'\Delta_d+\left\{\mathcal{H}(1-3w)-\frac{w}{1+w}\frac{\rho_d'}{\rho_d}+\frac{a^2Q_d^0}{\rho_d}\right \}\times\left\{-3\mathcal{H}C_e^2+3w\mathcal{H}-\frac{a^2Q_d^0}{\rho_d}\right\}\Delta_d\nonumber\\
&&-[\mathcal{H}+3\mathcal{H}C_e^2-6w\mathcal{H}+\frac{2a^2Q_d^0}{\rho_d}-\frac{w}{1+w}\frac{\rho_d'}{\rho_d}]\Delta_d'-k(\frac{a^2\delta
Q_{pd}^{I}}{\rho_d})-k^2C_e^2\Delta_d\nonumber\\
&&+3(w'\mathcal{H}+w\mathcal{H}')\Delta_d-k^2(1+w)\Psi+
\frac{a^2\delta
Q_{d}^{0I}}{\rho_d}[\mathcal{H}(1-3w)-\frac{w}{1+w}\frac{\rho_d'}{\rho_d}+\frac{a^2Q_d^0}{\rho_d}]+(\frac{a^2\delta
Q_{d}^{0I}}{\rho_d})'.
\end{eqnarray}
Changing the conformal time into the cosmic proper time, we can
rewrite the above second order equations as
\begin{eqnarray}
&&\ddot{\Delta}_m+2(H+\frac{aQ_m^0}{\rho_m})\dot{\Delta}_m+\left[\frac{aQ_m^0}{\rho_m}H+(\frac{aQ_m^0}{\rho_m})^2+\frac{1}{a}\dot{(\frac{a^2Q_m^0}{\rho_m})}\right]\Delta_m=\nonumber
\\
&&(H+\frac{aQ_m^0}{\rho_m})\frac{a\delta
Q_m^{0I}}{\rho_m}+\frac{1}{a}\dot{(\frac{a^2\delta
Q_m^{0I}}{\rho_m})}-k(\frac{\delta
Q_{pm}^{I}}{\rho_m})-\frac{k^2}{a^2}\Psi,
\end{eqnarray}
\begin{eqnarray}
&&\ddot{\Delta}_d+\left\{2H+3HC_e^2-6wH+\frac{2aQ_d^0}{\rho_d}-\frac{w}{1+w}\frac{\dot{\rho}_d}{\rho_d}\right\}\dot{\Delta}_d=-3(\dot{H}+H^2)C_e^2\Delta_d-\frac{1}{a}\dot{(\frac{a^2Q_d^0}{\rho_d})}\Delta_d\nonumber\\
&&+\left\{H(1-3w)-\frac{w}{1+w}\frac{\dot{\rho}_d}{\rho_d}+\frac{aQ_d^0}{\rho_d}\right\}\times\left\{-3HC_e^2+3wH-\frac{aQ_d^0}{\rho_d}\right\}\Delta_d-k(\frac{\delta
Q_{pd}^{I}}{\rho_d})-\frac{k^2}{a^2}C_e^2\Delta_d-k^2(1+w)\frac{\Psi}{a^2}\nonumber
\\
&&+3\left[\dot{w}H+w(\dot{H}+H^2)\right]\Delta_d+\frac{a\delta
Q_{d}^{0I}}{\rho_d}\left\{H(1-3w)-\frac{w}{1+w}\frac{\dot{\rho}_d}{\rho_d}+\frac{aQ_d^0}{\rho_d}\right\}+\frac{1}{a}\dot{(\frac{a^2\delta
Q_d^{0I}}{\rho_d})}.\nonumber
\end{eqnarray}
where the dot denotes the derivative with respect to the proper time
$t$. For the convenience in the following discussions, we can
further rewrite the second order perturbation equations for DE and
DM into dimensionless form
\begin{eqnarray}
&&\frac{d^2ln\Delta_m}{dlna^2}+\left[\frac{1}{2}-\frac{3}{2}w(1-\Omega_m)\right]\frac{dln\Delta_m}{dlna}+\left(\frac{dln\Delta_m}{dlna}\right)^2+\frac{1}{H^2}\left\{\frac{aQ_m^0}{\rho_m}H+\left(\frac{aQ^0_m}{\rho_m}\right)^2+\frac{H}{a}\frac{d}{dlna}\left(\frac{a^2Q_m^0}{\rho_m}\right)\right\}\nonumber\\
&=&\left(H+\frac{aQ_m^0}{\rho_m}\right)\frac{a\delta
Q_m^{0I}}{H^2\rho_m}e^{-ln\Delta_m}-\frac{2}{H}\frac{aQ_m^0}{\rho_m}\frac{dln\Delta_m}{dlna}+\frac{1}{aH}\frac{d}{dlna}\left(\frac{a\delta
Q_m^{0I}}{\rho_m}\right)e^{-ln\Delta_m}-\frac{k\delta
Q_{pm}^{I}}{H^2\rho_m}e^{-ln\Delta_m}\nonumber\\
&&+\frac{3}{2}\left[\Omega_m\Delta_m+(1-\Omega_m)\Delta_d\right]e^{-ln\Delta_m},
\end{eqnarray}
\begin{eqnarray}
&&\frac{d^2ln\Delta_d}{dlna^2}+\left(\frac{dln\Delta_d}{dlna}\right)^2+\left[\frac{1}{2}-\frac{3}{2}w(1-\Omega_m)\right]\frac{dln\Delta_d}{dlna}+\left(3C_e^2-6w+\frac{2aQ_d^0}{H\rho_d}-\frac{w}{1+w}\frac{1}{\rho_d}\frac{d\rho_d}{dlna}\right)\frac{dln\Delta_d}{dlna}\nonumber\\
&=&-3\left(\frac{1}{H}\frac{dH}{dlna}+1\right)C_e^2-\frac{1}{aH}\frac{d}{dlna}\left(\frac{a^2Q_d^0}{\rho_d}\right)+\left\{1-3w-\frac{w}{1+w}\frac{1}{\rho_d}\frac{d\rho_d}{dlna}+\frac{a}{H}\frac{Q_d^0}{\rho_d}\right\}\times\left\{-3C_e^2+3w-\frac{a}{H}\frac{Q_d^0}{\rho_d}\right\}\nonumber\\
&&-\frac{k\delta
Q_{pd}^{I}}{\rho_dH^2}e^{-ln\Delta_d}-\frac{k^2C_e^2}{a^2H^2}+(1+w)\frac{3}{2}\left\{\Omega_m\Delta_m+(1-\Omega_m)\Delta_d\right\}e^{-ln\Delta_d}+3\left[\frac{dw}{dlna}+w\left(\frac{1}{H}\frac{dH}{dlna}+1\right)\right]\nonumber
\\
&&+\frac{a\delta
Q_d^{0I}}{H\rho_d}\left\{1-3w-\frac{w}{1+w}\frac{1}{\rho_d}\frac{d\rho_d}{dlna}+\frac{aQ_d^0}{H\rho_d}\right\}e^{-ln\Delta_d}+\frac{1}{aH}\frac{d}{dlna}\left(\frac{a^2\delta
Q_d^{0I}}{\rho_d}\right)e^{-ln\Delta_d}.
\end{eqnarray}

In the subhorizon approximation, from the perturbed Einstein
equations eq(~\ref{Einstein}) we can get the ``Poission equation"
\begin{equation}
-\frac{k^2}{a^2}\Psi=\frac{3}{2}H^2\left\{\Omega_m\Delta_m+(1-\Omega_m)\Delta_d\right\}
\end{equation}
where we have used Friedmann equation in the derivation. This
equation can be used to build the bridge between the matter
perturbations to the metric perturbations.

Since the nature of DE and DM remains unknown, it will not be
possible to derive the precise form of the interaction between them
from the first principle. One has to assume a specific coupling from
the outset \cite{91,30,31} or determine it from phenomenological
requirements \cite{92,23}. From eq(\ref{Qtransform})and
eq(\ref{Lie}), we know that $Q^{\nu}$ is a covariant vector, which
does not need to depend on the four velocity. For the generality, we
can assume the phenomenological description of the interaction
between dark sectors in the comoving frame as
\begin{eqnarray}
Q_m^{\nu}=\left[\frac{3\mathcal{H}}{a^2}(\delta_1\rho_m+\delta_2\rho_d),0,0,0\right]^{T}\nonumber\\
Q_d^{\nu}=\left[-\frac{3\mathcal{H}}{a^2}(\delta_1\rho_m+\delta_2\rho_d),0,0,0\right]^{T},
\end{eqnarray}
where $\delta_1,\delta_2$ are small positive dimensionless constants
and superindex $T$ is the transpose of the vector. Choosing positive
sign in the interaction, one can ensure the direction of energy
transfer from DE to DM, which is required to alleviate the
coincidence problem \cite{he15,binp} and avoid some unphysical
problems such as negative DE density etc \cite{23,maartens}. In the
subhorizon approximation $k>>aH$,
\begin{eqnarray}
\delta Q^{0I}_{m}&\simeq&
\frac{3\mathcal{H}}{a^2}(\delta_1\delta\rho_m^I+\delta_2\delta\rho_d^I)\nonumber\\
\delta Q^{0I}_{d}&\simeq&
\frac{3\mathcal{H}}{a^2}(\delta_1\delta\rho_m^I+\delta_2\delta\rho_d^I)\nonumber\\
\delta\rho_m^I&\simeq& \delta \rho_m\nonumber \\
\delta\rho_d^I&\simeq& \delta \rho_d.\nonumber \\
\end{eqnarray}
Further noting that the gauge-invariant momentum transfer $\delta
Q_{p\lambda}^{I}$ refers to the intrinsic momentum transfer between
dark sectors and as explained in \cite{Sa}, such intrinsic momentum
transfer is due to the collision of particles from different fluids.
Such collision can produce acoustics in DM fluid as well as pressure
which may resist the squeeze of the attraction of gravity and hinder
the growth of gravity fluctuations during tightly coupled photon
baryon period, we therefore set $\delta Q_{pm}^{I}\approx\delta
Q_{pd}^{I}\approx0$\cite{he}.This is a choice of interaction and the
result should not heavily depend on such setting.

Employing the above interaction form, we can finally arrive at
perturbation equations for dark sectors with constant EoS of DE
\begin{eqnarray}
&&\frac{d^2ln \Delta_m}{dlna^2}  = -\left( \frac{dln \Delta_m }{d
lna }\right)^2 - \left [ \frac{1}{2} - \frac{3}{2}w (1 -
\Omega_m)\right] \frac{dln \Delta_m}{d lna}
 -(3\delta_1 +
6\frac{\delta_2}{r})\frac{dln
\Delta_m}{dlna}+3\frac{\delta_2}{r}\frac{dln \Delta_d}{dlna} exp(ln
\frac{\Delta _d}{ \Delta _m})\nonumber\\
&&+\frac{3[exp (ln \frac{\Delta _d}{ \Delta
_m})-1]}{r}\left(\delta_2+3\delta_1\delta_2+3\delta_2^2/r+\delta_2(\frac{H'}{H}+1)-
\delta_2\frac{r'}{r}\right)+\frac{3}{2}\left[\Omega_m+(1 -
\Omega_m)exp (ln \frac{\Delta _d}{ \Delta
_m})\right]\label{DMevolution}
\end{eqnarray}
\begin{eqnarray}
&&\frac{d^2ln\Delta_d}{dlna^2}= -\left(\frac{dln
\Delta_d}{dlna}\right)^2 - \left[\frac{1}{2} -
\frac{3}{2}w(1-\Omega_m)\right]\frac{dln \Delta_d}{d lna} + (1 +
w)\frac{3}{2}\left[\Omega_mexp(ln\frac{\Delta_m}{\Delta_d})+
(1 - \Omega_m)\right]-\frac{k^2C_e^2}{a^2H^2}\nonumber \\
&+&\left[3\delta_2+6\delta_1r+6w-3C_a^2+3(C_e^2-C_a^2)\frac{\delta_1r+\delta_2}{1+w}+\frac{C_e^2}{1+w}\frac{\rho_d'}{\rho_d}\right]\frac{dln\Delta_d}{dlna}\nonumber
\\
&-&3\delta_1 r
exp(ln\frac{\Delta_m}{\Delta_d})\frac{dln\Delta_m}{dlna}+3(\frac{H'}{H}+1)(w-C_e^2)
+3\delta_1(\frac{H'}{H}r+r+r')- 3\delta_1\left[(\frac{H'}{H}+1)r+r'\right]exp(ln\frac{\Delta_m}{\Delta_d})\nonumber \\
&+&3\left[w - C_e^2 + \delta_1r(1-
exp(ln\frac{\Delta_m}{\Delta_d}))\right]\left[(1-3w)-3\frac{C_e^2-C_a^2}{1+w}(1+w+\delta_1r+\delta_2)-3(\delta_1r+\delta_2)-\frac{C_e^2}{1+w}\frac{\rho_d'}{\rho_d}\right],\label{DEevolution}
\end{eqnarray}
where $r=\rho_m/\rho_d$ and the prime denotes $d/d ln a$.

\section{numerical results}

In this section we present numerical results of solving the above
perturbation equations. We concentrate on the behavior of the
evolution of the DM perturbation.

\begin{figure}
\includegraphics[width=3.5in,height=3.5in]{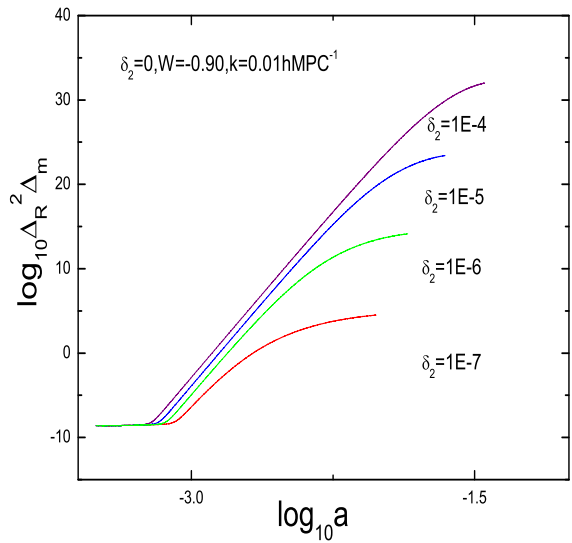}
\caption{This figure shows the blow up of the DM perturbation when
the coupling between dark sectors is proportional to the energy
density of DM. $\Delta_R^2=2.41\times10^{-9}$ is the amplitude of
curvature perturbations from WMAP five-year results and $\Delta_m$
is the density contrast of DM.}\label{blowup}
\end{figure}

It was argued that the presence of the interaction between DE and DM
may give rise to dynamical instabilities in the growth of structure
\cite{33}. In our general form of the phenomenological interaction,
when we choose the coupling between dark sectors in proportion to
the DM energy density by setting $\delta_2=0$ while keeping
$\delta_1$ nonzero, we do find the consistent fast growth of the
fluctuations of DM, as seeing in Fig ~\ref{blowup}. However, when we
choose the dark sectors' mutual interaction in proportion to the
energy density of DE by taking $\delta_1=0$ and $\delta_2\neq 0$, we
have the stable DM perturbation. This result tells us that the
stability in the growth of structure depends on the type of coupling
between dark sectors, which is consistent with the findings in the
curvature perturbation in \cite{he}.

\begin{figure}
\begin{center}
  \begin{tabular}{cc}
\includegraphics[width=2.8in,height=2.8in]{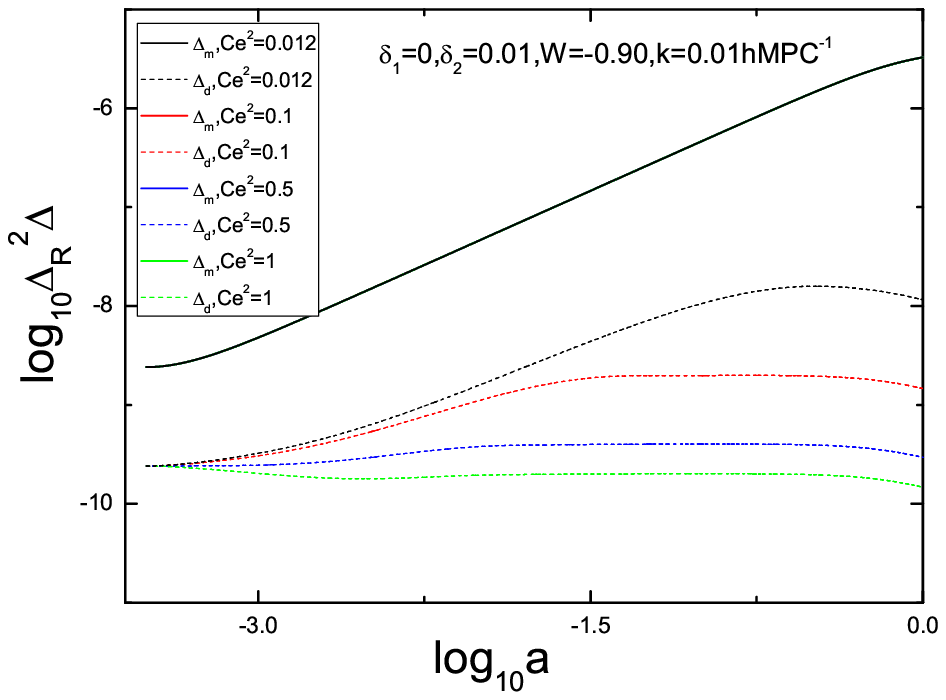}&
\includegraphics[width=2.8in,height=2.8in]{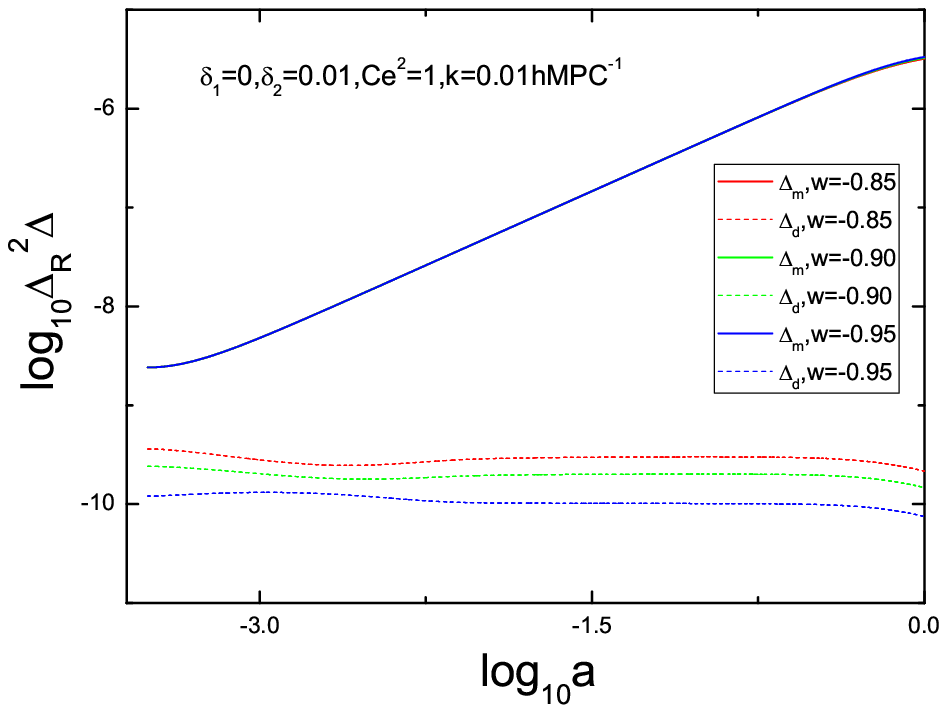}\nonumber \\
    (a)&(b)\nonumber \\
\includegraphics[width=2.8in,height=2.8in]{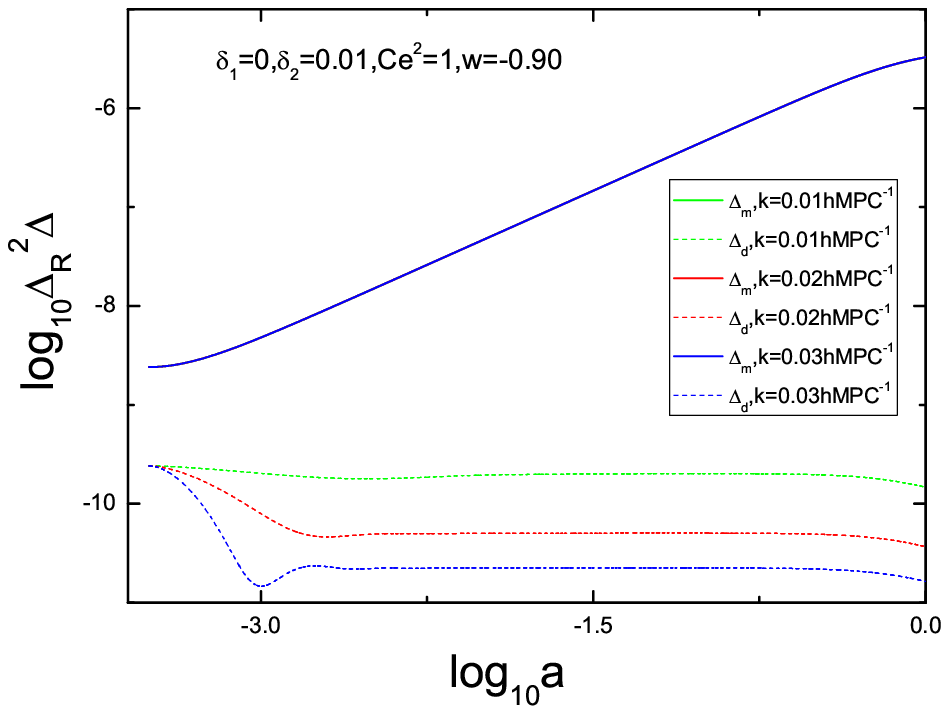}&
\includegraphics[width=2.8in,height=2.8in]{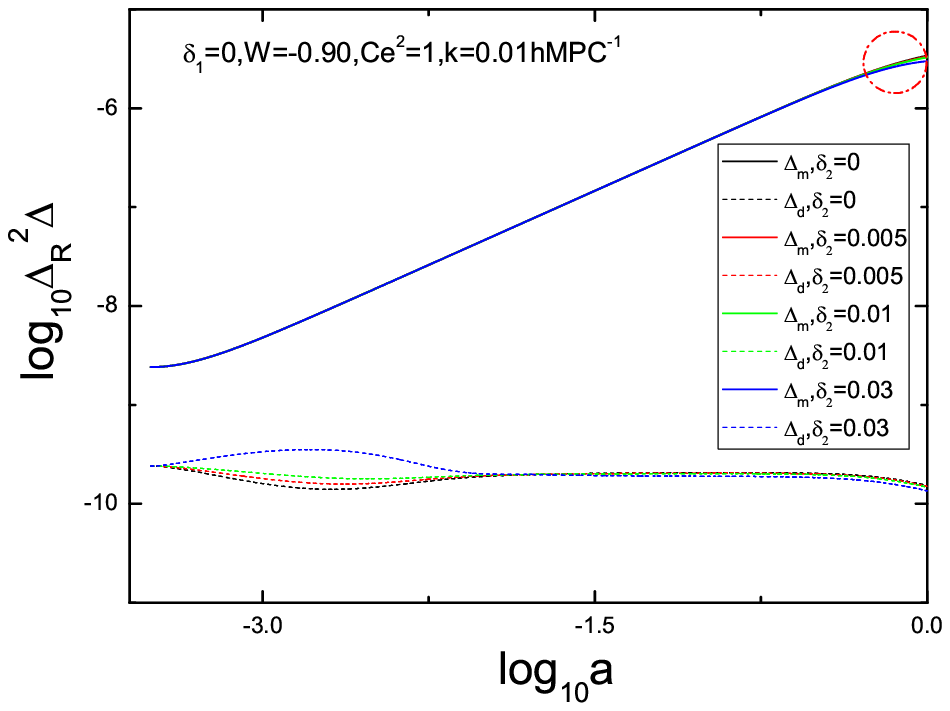}\nonumber \\
    (c)&(d)\nonumber
  \end{tabular}
\end{center}
\caption{These figures illustrate the behaviors of dark
fluctuations in different cases when varying the effective sound
speed,dark energy EoS,wave number,and coupling. The solid lines
are for the DM perturbation while the dotted lines are for the DE
perturbation. The solid lines in each panel are largely
overlapped, except for that part circled in panel
(d)}\label{fluctuation}
\end{figure}

In the following we focus on the interaction proportional to the
energy density of DE to keep the stability of the matter density
perturbation. Initial conditions are given at the redshift $z=3200$
where approximately is the time of matter-radiation equality, and we
take the adiabatic initial conditions and assume zero initial time
derivatives of DM and DE perturbations. In our analysis we do not
allow $w<-1$. As to $k$, we choose its value above $0.01h\rm
Mpc^{-1}$ so that there is large scale structure data on the matter
power spectrum \cite{8,3}. For the sound speed of DE, $C_e^2$, we
restrict it to be positive and smaller than unity. In
Fig~\ref{fluctuation}, we show the evolution of perturbations of DM
and DE for different values of parameters $C_e^2,w,k$ and the
strength of the coupling $\delta_2$ between dark sectors.

Without the interaction between dark sectors, we observed that there
is the sensible influence of the DE perturbation on the evolution of
DM perturbation when the sound speed is tiny enough and the EoS of
DE substantially deviates from $-1$. In Fig~\ref{fluctuation} a, we
can see this qualitative behavior that for smaller $C_e^2$,the DE
perturbation grows. If we take $w$ further away from $-1$, it will
grow more and will influence the DM perturbation. In
Fig~\ref{fluctuation} b, it shows that for fixed $C_e^2$, DE
perturbation grows when $w$ deviates from $-1$. This is consistent
with the result in \cite{8}. However, when $C_e^2$ is not so tiny
and $w$ close to $-1$, in the subhorizon approximation, we observed
that the influence of the DE perturbation is suppressed. For bigger
$k$, DE influence is even smaller, see Fig~\ref{fluctuation} c. This
result will not change when the interaction presents. The influence
of the interaction between dark sectors can start to appear in the
very recent epoch. In Fig~\ref{fluctuation} d, it shows there in the
red circle the small deviation caused by the coupling between dark
sectors in the matter density perturbation at recent time.
\begin{figure}
\begin{center}
  \begin{tabular}{cc}
\includegraphics[width=2.8in,height=2.8in]{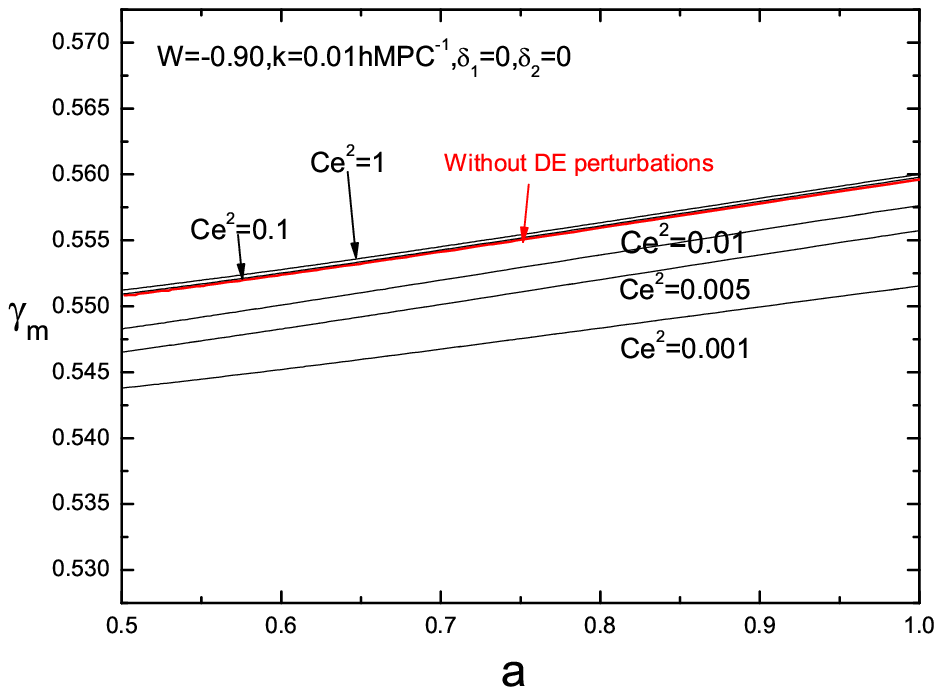}&
\includegraphics[width=2.8in,height=2.8in]{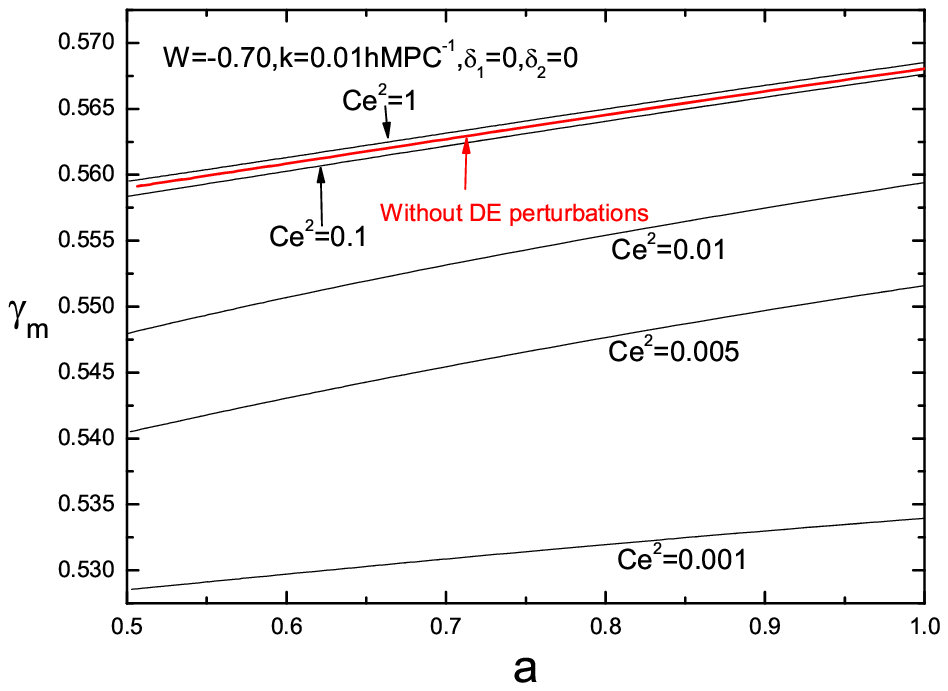}\nonumber \\
    (a)&(b)\nonumber \\
\includegraphics[width=2.8in,height=2.8in]{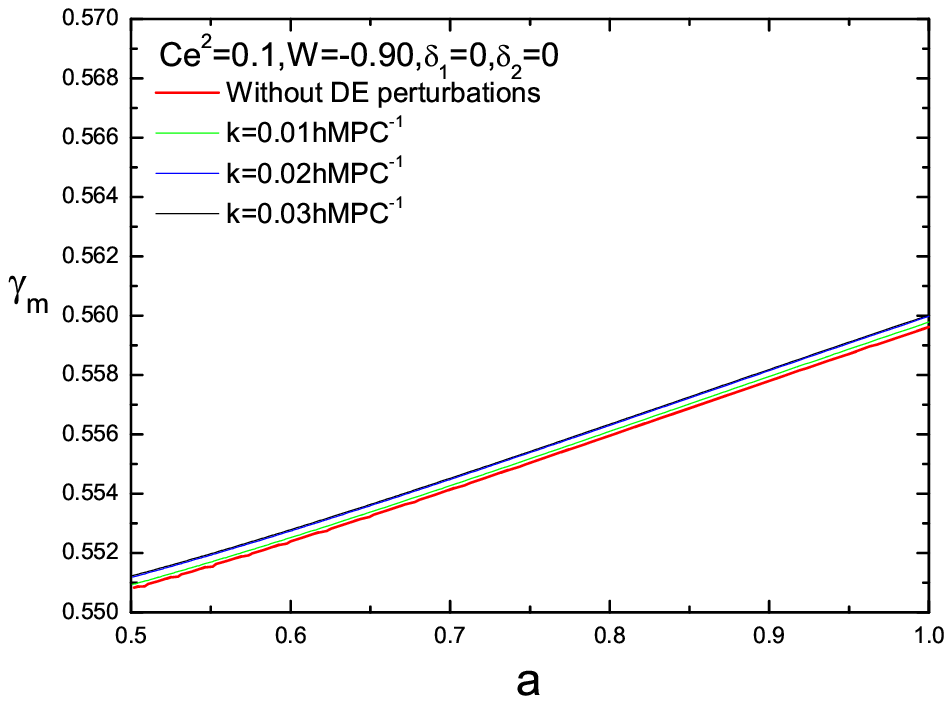}&
\includegraphics[width=2.8in,height=2.8in]{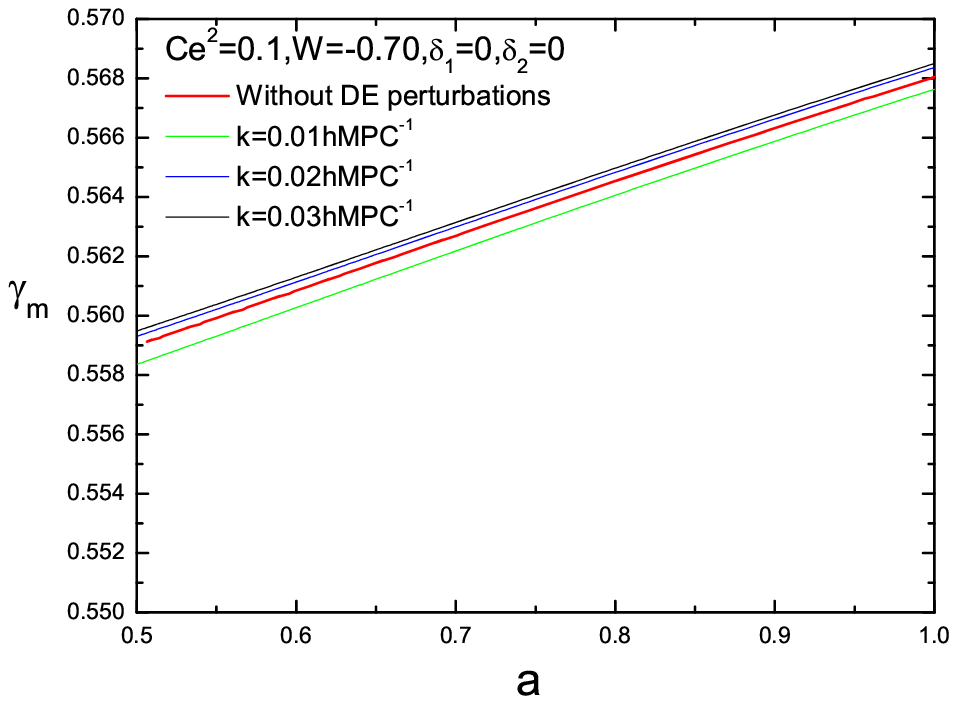}\nonumber \\
    (c)&(d)\nonumber
  \end{tabular}
\end{center}
\caption{These figures illustrate the behaviors of dark fluctuations
in different cases when varying the effective sound speed,dark
energy EoS,wave number,and coupling. The solid lines are for the DM
perturbation while the dotted lines are for the DE
perturbation.}\label{growthindex_without_coupling}
\end{figure}

To see more clearly of the influence of different parameters on
the growth history of the DM perturbation, we introduce the growth
index $\gamma$ with the definition \cite{8}
\begin{equation}
\gamma_m=(\ln
\Omega_m)^{-1}\ln\left(\frac{a}{\Delta_m}\frac{d\Delta_m}{da}\right).
\end{equation}
The growth index is generically not constant which was first
emphasized and investigated in terms of cosmological parameters in
\cite{pp}.  The growth index has been argued as a useful way in
principle to distinguish the modified gravity models from DE models
\cite{7,p}. In Fig~\ref{growthindex_without_coupling}, we show that
the influence of the DE perturbation on the growth index for the
lack of interaction between dark sectors. The red lines in the
figure mark the result without DE perturbation. In
Fig~\ref{growthindex_without_coupling} a, we see that for fixed DE
EoS $w$, when $C_e^2$ decreases, the growth index with the DE
perturbation deviates more from the result without DE perturbation.
This deviation will be even more prominent when DE EoS is further
away from $-1$ as shown in Fig.~\ref{growthindex_without_coupling}
b. We can see clearly that the difference between the growth index
with and without DE perturbation can be as big as $0.03$. Our
numerical result further supports what found in \cite{8}. In
Fig~\ref{growthindex_without_coupling} c,d, we observed that in the
subhorizon approximation the value of $k$ does not influence as much
as the parameters $w, C_e^2$ on the result of the growth index.

\begin{figure}
\includegraphics[width=2.8in,height=2.8in]{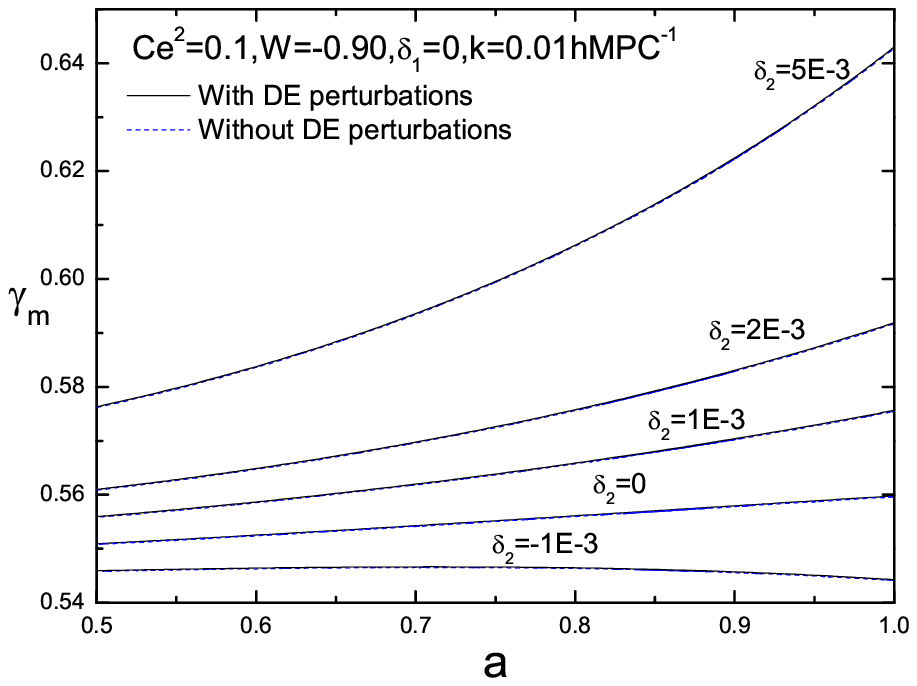}
\includegraphics[width=2.8in,height=2.8in]{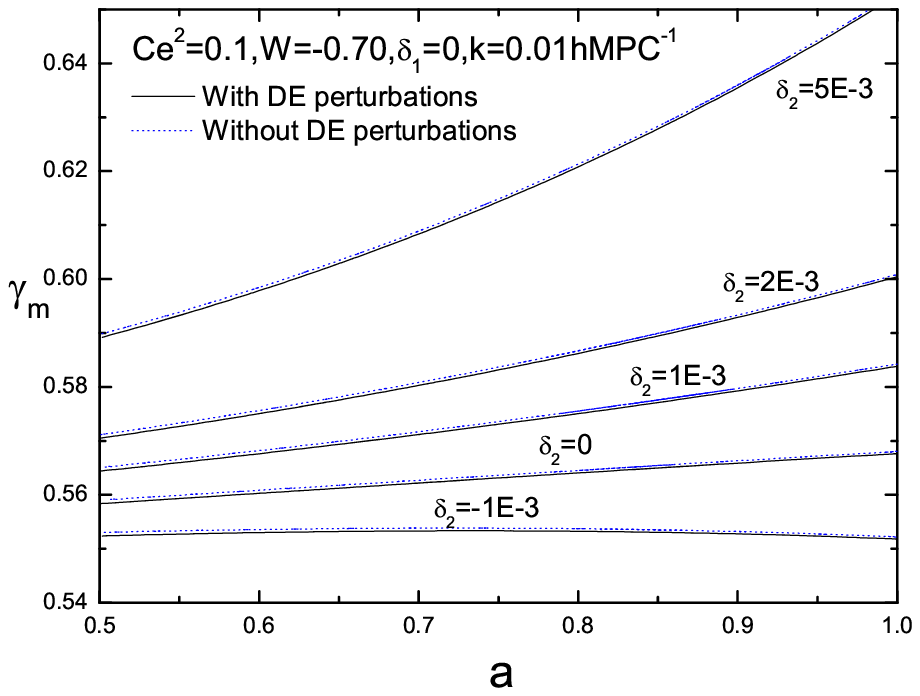}
\caption{The growth index behavior when the interaction between DE
and DM presents. Solid lines are for the result with DE
perturbation, while dotted lines are for the result without DE
perturbation.}\label{couple_growth_index}
\end{figure}

In Fig~\ref{couple_growth_index}, we present our numerical results
when we incorporate the interaction between DE and DM. The solid
lines are for the results with DE perturbation, while the dotted
lines are for the results without DE perturbation. It is clearly
shown that the growth index got more influenced from the interaction
between dark sectors than the DE perturbation. If we take the best
fitting value of $\delta_2$ from observations of the expansion
history of the universe, $\delta_2\sim 10^{-2}$ \cite{23,24}, its
influence on the growth index will even overwhelm that of the DE
perturbation. Although the enhancement of the growth index due to
the DE perturbation and the interaction shown in
Fig.~\ref{growthindex_without_coupling},~\ref{couple_growth_index}
is clear, the available accuracy from the observations such as
DUNE\cite{riotto32} etc. is calculated for the $\Lambda$CDM model
and may not be true for our interacting cosmology, since the current
values of $\Omega_{c0}, \delta_{c0}$ may not typically be equal in
the two cases as argued in \cite{34}. However, this phenomenon is
interesting, as it opens the possibility that the future measurement
of the growth factor may be helpful to reveal the presence of the
interaction between DE and DM.

\section{conclusions and discussions}

The cosmological observations have provided a firm evidence for
significant physics beyond standard models. It is clear now that the
formation of structure in the universe demands DM and the
accelerated expansion of our universe requires some kind of DE or a
significant infrared modification on Einstein gravity. DE and DM are
two major components of the cosmic energy budget. It is reasonable
to explore possible interactions between them in the framework of
field theory.

In this paper we have concentrated on the time evolution of the DM
perturbation. We have derived general equations to describe the
perturbations of DM and DE which incorporates the interactions
between them. It was argued that the interaction between dark
sectors might give rise to dynamical instabilities on the growth of
structure \cite{33}. However, we observed that this instability
depends on the form of the coupling. For example, if we choose the
interaction in proportion to the energy density of DE, we did
observe the stable growth of structure. This result is consistent
with what is found in the curvature perturbation \cite{he}.

Besides the interaction between dark sectors, we have also discussed
the effects of the nonvanishing DE perturbation on the evolution of
DM perturbation. Usually in the discussion of the growth of
structure, the DE perturbation was neglected. For the minimal
picture without the coupling between dark sectors, the influence of
the DE perturbation on the growth function of DM perturbation has
been examined in \cite{8}. In our general formalism, by taking the
appropriate coupling between dark sectors which enables the stable
growth of structure, we have found that the effect of the
interaction between dark sectors overwhelms that of the DE
perturbation on the growth function of DM perturbation. When the DE
EoS $w$ is in the vicinity of $-1$ somewhere abound the best fitted
value at the moment, the DE perturbation is suppressed, however,
when the interaction presents, the growth index can differ from the
value without interaction by a big amount up to the observational
sensibility. This provides an interesting way to probe the
interaction between dark sectors through the observations on the
growth of structure in large scale \cite{23,24}.

It would be of great interest to confront our theoretical work to
observations to constrain the interaction between dark sectors.
However, the data available on the growth of structure are still
poor and there is still a long way to go before we can talk about
precision cosmology in this respect. We will leave our work in this
direction in the future. On the other hand, it is also very
interesting to include the interaction between dark sectors to
modify the code in studying the N-body cosmological simulations on
the structure formation. Work in this direction is in progress.

\acknowledgments{This work has been supported partially by NNSF of
China, Shanghai Science and Technology Commission and Shanghai
Education Commission.}


\begin{thebibliography}{99}
\bibitem{1} S. J. Perlmutter et al., Nature 391, 51 (1998); A. G. Riess et al., Astron. J. 116, 1009 (1998);
S. J. Perlmutter et al., Astroph. J. 517, 565 (1999); J. L. Tonry
et al., Astroph. J. 594, 1, (2003);  A. G. Riess et al., Astroph.
J. 607, 665 (2005); P. Astier et al., Astron. Astroph. 447, 31
(2005); A G. Riess et al., Astroph. J. 659, 98 (2007).

\bibitem{2} M. Kowalski et al., 2008, arXiv:0804.4142.

\bibitem{3}M. Tegmark et al., Phys. Rev. D74, 123507 (2006).

\bibitem{4} S. M. Carroll, V. Duvvuri, M. Trodden, and M. S. Turner, Phys. Rev. D70, 043528
(2004).

\bibitem{5} G. R. Dvali, G. Gabadadze, and M. Porrati, Phys. Lett. B485, 208
(2000); A. Sheykhi, B. Wang, and N. Riazi, Phys. Rev. D 75, 123513
(2007);  S. Y. Yin, B. Wang, E. Abdalla, C. Y. Lin, Phys.Rev.D76,
124026, (2007).

\bibitem{6} D. N. Spergel et al. (WMAP), Astrophys. J. Suppl. 148,
175 (2003), astro-ph/0302209; J. L. Sievers et al. (2005),
astro-ph/0509203;  D. N. Spergel et al. (WMAP), Astrophys. J.
Suppl. 170, 377 (2007), astro-ph/0603449;  C.-L. Kuo et al.
(2006), astro-ph/0611198;  C. L. Reichardt et al. (2008),
0801.1491;  M. R. Nolta et al. (WMAP) (2008), 0803.0593; S. Cole
et al. (The 2dFGRS), Mon. Not. Roy. Astron. Soc. 362, 505 (2005),
astro-ph/0501174;  M. Tegmark et al., Phys. Rev. D74, 123507
(2006), astroph/ 0608632; W. J. Percival et al., Astrophys. J.
657, 645 (2007), astroph/ 0608636;  D. J. Eisenstein et al.
(SDSS), Astrophys. J. 633, 560 (2005), astro-ph/0501171;  W. J.
Percival et al., Mon. Not. Roy. Astron. Soc. 381, 1053 (2007),
0705.3323; A. G. Riess et al. (Supernova Search Team), Astron. J.
116, 1009 (1998), astro-ph/9805201; S. Perlmutter et al.
(Supernova Cosmology Project), Astrophys. J. 517, 565 (1999),
astro-ph/9812133; P. Astier et al., Astron. Astrophys. 447, 31
(2006), astroph/ 0510447; A. G. Riess et al., Astrophys. J. 659,
98 (2006), astroph/ 0611572.

\bibitem{7} E. Bertschinger and P. Zukin, arXiv:0801.2431; S. Tsujikawa, Phys. Rev. D76,
023514, (2007);  B. Jain and P. Zhang, ArXiv: 0709.2375;  H. Wei
and S. N. Zhang, ArXiv: 0803.3292; Y. G. Gong, ArXiv: 0808.1316.

\bibitem{8}  G. Ballesteros, A. Riotto, Phys. Lett. B 668,
171 (2008).

\bibitem{zhang}  G. Zhao, J. Xia, H. Li, C. Tao, J. Virey, Z. Zhu, X.M.
Zhang, Phys. Lett. B648, 8,(2007).


\bibitem{9} L. Amendola, Phys. Rev. D62, 043511 (2000); L. Amendola and C. Quercellini, Phys. Rev. D68, 023514 (2003); L.
Amendola, S. Tsujikawa and M. Sami, Phys. Lett. B632, 155 (2006).

\bibitem{91} D. Pavon, W. Zimdahl, Phys. Lett. B628, 206 (2005), S. Campo, R. Herrera, D.
Pavon, Phys. Rev.D 78, 021302(R) (2008).

\bibitem{92} G. Olivares, F. Atrio-Barandela and D. Pavon, Phys. Rev.
D74, 043521 (2006).

\bibitem{93} C. G. Boehmer, G. Caldera-Cabral, R. Lazkoz, R. Maartens,
Phys. Rev. D 78 (2008) 023505.

\bibitem{99} S. B. Chen, B. Wang, J. L. Jing, Phys.Rev.D78, 123503
(2008).

\bibitem{10} B. Wang, J. Zang, C.-Y. Lin, E. Abdalla and S. Micheletti, Nucl. Phys. B778, 69 (2007).

\bibitem{11} W. Zimdahl, Int. J. Mod. Phys. D14, 2319 (2005).

\bibitem{binp} D. Pavon, B. Wang, Gen. Relav. Grav. (in press),
arXiv:0712.0565.

\bibitem{98}  B. Wang, C.-Y. Lin, D. Pavon, E. Abdalla, Phys.Lett.B 662, 1,
(2008).


\bibitem{13} O. Bertolami, F. Gil Pedro and M. Le Delliou, Phys. Lett. B654, 165 (2007). O. Bertolami, F. Gil Pedro and M. Le Delliou,
arXiv:0705.3118v1.


\bibitem{16} Z. K. Guo, N. Ohta and S. Tsujikawa, Phys. Rev. D76, 023508 (2007).

\bibitem{maartens} J. Valiviita, E. Majerotto, R. Maartens, JCAP 07,(2008)020, arXiv:0804.0232.

\bibitem{he}  J. H. He, B. Wang, E. Abdalla, Phys. Lett. B 671, 139 (2009), arXiv:0807.3471


\bibitem{he15} W. Zimdahl, D. Pavon, L.P. Chimento, Phys. Lett. B 521 (2001) 133; L.P. Chimento, A.S. Jakubi, D. Pavon, W. Zimdahl,
Phys. Rev. D 67 (2003) 083513.


\bibitem{22} E. Abdalla, L.Raul W. Abramo, L. Sodre Jr., B. Wang, arXiv:0710.1198 [astro-ph].

\bibitem{23} J. H. He, B. Wang, JCAP 06, 010 (2008), arXiv:0801.4233.

\bibitem{24} C. Feng, B. Wang, E. Abdalla, R. K. Su, Phys. Lett. B665, 111 (2008), arXiv:0804.0110.

\bibitem{new}  B. Jackson, A. Taylor, A. Berera, arXiv:0901.3272.

\bibitem{riotto32} A. Refrefier, et al, ArXiv: 0802.2522.

\bibitem{30} S. Das, P. S. Corasaniti and J. Khoury, Phys. Rev. D73, 083509 (2006).

\bibitem{31} L. Amendola, D. Tocchini-Valentini, Phys. Rev. D 64 (2001) 043509; G. W. Anderson, S. M. Carroll, astro-ph/9711288.

\bibitem{Sa} V. Mukhanov, Physical Foundation of Cosmology,
Cambridge Univ. Press, 2005; H. Kodama, M. Sasaki, Prog. Theor.
Phys. Suppl. 78, 1, (1984).

\bibitem{33} N. Afshordi, M. Zaldarriaga, and K. Kohri, Phys. Rev.D72, 065024 (2005),
astro-ph/0506663; M. Kaplinghat and A. Rajaraman, Phys. Rev. D75,
103504 (2007), astro-ph/0601517;O. E. Bjaelde et al., JCAP 0801,
026 (2008), 0705.2018;R. Bean, E. E. Flanagan, and M. Trodden, New
J. Phys. 10, 033006 (2008), 0709.1124;R. Bean, E. E. Flanagan, and
M. Trodden, Phys. Rev. D78, 023009 (2008), 0709.1128;L. Vergani,
L. P. L. Colombo, G. La Vacca, and S. A. Bonometto (2008),
0804.0285.

\bibitem{34} G. Caldera-Cabral, R. Maartens, B. M.
Schaefer,ArXiv: 0905.0492.

\bibitem{pp} D. Polarski, R. Gannouji
Phys. Lett. B660, 439 (2008).

\bibitem{p} R. Gannouji, D. Polarski
JCAP 0805, 018 (2008); R. Gannouji, B. Moraes, D. Polarski
arXiv:0809.3374.





\end{thebibliography}
\end{document}